\documentclass{aa}  
\usepackage{graphicx}
\usepackage{lipsum}
\usepackage{txfonts}

\begin{document}

    \title{Lithium enrichment on the single active K1-giant DI\,Piscium \thanks{Data were gathered via OPTICON FP7 in semester 2012B.}} 
  \subtitle{Possible joint origin of differential rotation and Li enrichment}

   \author{L. Kriskovics
          \inst{1},
          Zs. K\H{o}v\'ari\inst{1},
	  K. Vida\inst{1},
	  T. Granzer\inst{2}
	\and
	  K. Ol\'ah\inst{1}
    }
   \institute{Konkoly Observatory of the Hungarian Academy of Sciences, Konkoly Thege út 15–17, 1121, Budapest, Hungary\\
              \email{kriskovics@konkoly.hu}
	\and
              Leibniz Institute for Astrophysics (AIP), An der Sternwarte 16, D-14482 Potsdam, Germany
             }

   \date{Received September 15, 1996; accepted March 16, 1997}

  \abstract
   {}
   {We investigate the surface spot activity  of the rapidly rotating, lithium-rich active single K-giant DI Psc to measure the surface differential rotation and understand the mechanisms behind the Li-enrichment.}
   {Doppler imaging was applied to recover the surface temperature distribution of DI Psc in two subsequent rotational cycles using the individual mapping lines Ca \,{\sc i}\,6439, Fe\,{\sc i}\,6430, Fe\,{\sc i}\,6421 and Li\,{\sc i}\,6708. Surface differential rotation was derived by cross-correlation of the subsequent maps. Difference maps are produced to study the uniformity of Li-enrichment on the surface. These maps are compared with the rotational modulation of the Li\,{\sc i}\,6708 line equivalent width. }
   {Doppler images obtained for the Ca and Fe mapping lines agree well and reveal strong polar spottedness, as well as cool features at lower latitudes. Cross-correlating the consecutive maps yields antisolar differential rotation with shear coefficient $\alpha=-0.083\pm0.021$. The difference of the average and the Li maps indicates that the lithium abundance is non-activity related. There is also a significant rotational modulation of the Li equivalent width.  }
 {}

   \keywords{stars: activity --
                stars: imaging --
                stars: late-type --
		stars: individual --
		starspots --
		stars: individual: DI\,Psc
               }
   \titlerunning{Lithium enrichment on the single active K1-giant DI\,Psc}            
   \authorrunning{Kriskovics et al.}
   \maketitle
%
\section{Introduction}
As a star evolves towards the red giant branch (RGB), it arrives at an evolutionary phase  on the ascent of the RGB. Here the convective envelope deepens, and the surface Li abundance is decreased as 
the convective motions carry the Li-rich material towards the inner, hotter regions, where the lithium is diluted and depleted. This evolutionary stage is called the first dredge-up phase. As a result of this phenomenon, the measured surface lithium abundance is expected to be $A(\mathrm{Li}) < 1.5$. However, a rather small set of stars in this evolutionary stage shows an unusually high surface Li abundance (up to $\approx 2.5$), which cannot be explained well by the current stellar evolutionary models. 

Events capable of explaining this elevated Li abundance follow two distinct directions. The first one concerns external phenomena as contamination of the upper layers of the star by accretion of material in the form of a companion, for example planets or brown dwarfs \citep{brown, gatton}, while the second one is related to internal processes.
Among the scenarios proposed to explain the elevated lithium abundance by internal courses, one of the most well-known is the \cite{cameron1} process.
In this case, ${}^7\mathrm{Be}$ is produced in the inner, hotter layers via the ${}^4\mathrm{He}(\alpha, \gamma){}^7\mathrm{Be}$ reaction. If an enhanced outward convective flow transports the freshly synthesised beryllium towards the cooler regions, the lithium procuced by the ${}^7\mathrm{Be}(e^-, \nu){}^7\mathrm{Li}$ process can be preserved. In RGB stars, this enhanced mixing can be explained by rapid rotation \citep{charbonnel1, drake1, lugano1}. 

The possible connection between rotation and Li surface abundance has been investigated by several authors. \cite{demedeiros1} found that giants later than G0 were generally slow rotators and exhibit lower Li abundance, while hotter stars were rotating faster and show more surface lithium. There is, however, a small group of cool giants with an unusually high $A(\mathrm{Li})$ and fast rotation (e. g. \citealt{fekel1, drake1}). \cite{charbonnel1} pointed out the existence of two distinct strips on the RGB where this additional mixing occurs, and mentioned a high rotational rate as one of the possible inducers.

Rapid rotation is another well-known cause of enhanced stellar activity. Based on the well-observed fact that the surface lithium abundance is higher in sunspots than in quiet regions, the measured Li equivalent widths of disk-integrated stellar spectra can be affected by spottedness \citep{giampapa1}. However, in hot plage regions the Li abundance is lower, which can counterbalance this effect. Thus, studies that aim to investigate the relation between activity and Li-enhancement are inconclusive (cf. \citealt{jeffries1, fernandez1, berdyugina1}). \cite{giampapa1}, \cite{fekel2} and others also pointed out that Li abundance measurements could be affected by non-uniform temperature distribution. Since the Li\,{\sc i}\,6708 line is highly sensitive to temperature changes in the temperature regime of K stars due to its low ionization potential, a higher equivalent width is expected for increased spottedness.

\begin{table}[t!!] 
\caption{Averaged spectra used in the two series. The table gives the mean phases and HJDs of the observations, and the calculated signal-to-noise ratios at 6480 \AA.}
\label{specs}
\centering 
\begin{tabular}{ccc|ccc}
& Series I & & &Series II  & \\
\hline\hline  
      $\phi_{\mathrm{rot}}$&HJD& S/N &$\phi_{\mathrm{rot}}$& HJD& S/N\\
& -2456000  & & & -2456000&   \\
\hline\hline
0.035 & 214.3238 & 466 & 0.039 & 232.3354 & 433\\
0.371 & 202.4162 & 565 & 0.095 & 251.2953 & 591 \\
0.421 & 203.3953 & 546 & 0.150 & 234.3355 & 469 \\
0.483 & 204.4186 & 535 & 0.151 & 252.2921 & 591 \\
0.539 & 205.4321 & 352 & 0.321 & 237.3939 & 579 \\
0.594 & 206.4202 & 301 & 0.375 & 238.3655 & 261 \\
0.872 & 229.3381 & 535 & 0.425 & 239.2611 & 331 \\
0.986 & 213.4494 & 700 & 0.595 & 224.3826 & 431 \\
&&&0.926 & 230.3121 & 238 \\ 
      
\hline\hline                           
\end{tabular}
\end{table}

\begin{table}[t!!] 
\caption{Astrophysical parameters of DI\,Psc used during the Doppler imaging process.}
\label{astdata}
\centering 
\begin{tabular}{c|c}
\hline\hline
Spectral type & K1 III \\
\hline\hline
$\log{g}$ (adopted) & 2.5 \\
$T_{\mathrm{eff}}$ & $4600\pm100$ \\
$v\sin{i}$ ($\mathrm{km}\,\mathrm{s}^{-1}$) &$42$ \\
Inclination (${}^{\circ}$) & $50\pm10$ \\
$P_{\mathrm{rot}}$ (days) & $18.02$ \\
LTE Li abundance (log) & $2.20\pm0.15$ \\
\hline\hline
\end{tabular}
\end{table}
In our recent paper (\citealt{kovari1}, hereafter Paper I), we investigated the possible connections between rapid rotation, enchanced stellar activity, and elevated Li abundance on DI\,Psc (HD\,217352) and DP\,CVn (HD\,109703), two fast-rotating Li-rich K1 giants identified by the Vienna-KPNO Doppler imaging candidate survey \citep{strassmeier1}. We concluded, in accordance with \cite{fekel1}, that these stars with unusually high Li-abundance ($\mathrm{A}_{\mathrm{DI\,Psc}}(\mathrm{Li})=2.20$, $\mathrm{A}_{\mathrm{DP\,CVn}}(\mathrm{Li})~=~2.28$) had just left their first dredge-up. The spot configuration of the stars was investigated by means of Doppler imaging and proved to be unspectacular in that there were no large polar features present. Antisolar differential rotation of DP\,CVn ($\alpha=-0.035\pm0.010$) was also detected. This finding may agree with the enhanced mixing, since angular momentum can also be transported towards the poles by this additional convection, resulting in an antisolar-type differential rotation.

Rapid rotation can also affect the length of dynamo cycles. \cite{saar1} found that a correlation can be established between rotational period and cycle length.
\cite{lindborg1} classified DI\,Psc as a member of the active branch. According to \cite{saar1}, stars on this branch systematically show activity cycles that are two orders of magnitude higher than their rotational periods, i.e. a few years. Based on the large variability in spottedness even on a scale of a few years, \cite{lindborg1} suggested that the activity cycle of DI\,Psc is shorter than 10 years.

In this paper, we perform a new Doppler imaging study of DI\,Psc. Surface differential rotation is investigated from the time evolution of the spot distribution, to verify or rule out the dependency of Li equivalent width on spottedness. The 
surface lithium distribution is also studied, as is the rotational modulation of the Li {\sc i} 6708 equivalent widths. 

\begin{figure}[t!]
\centering
\includegraphics[width=9.2cm]{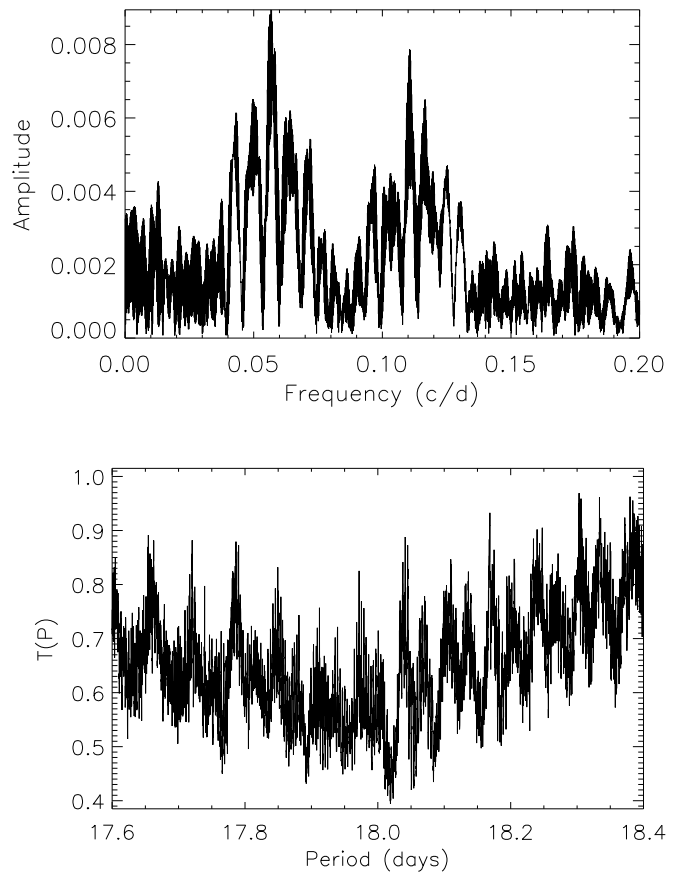}
\caption{Top panel: power spectrum of all avaible {\it V} and {\it y} data. The two peaks at 0.0561 and 0.1098 cycles/day correspond to  $P_{\mathrm{rot}}=17.82\,\mathrm{d}$ and $P_{\mathrm{rot}}=9.11\,\mathrm{d}$, respectively. Bottom panel: result of the SLLK fit (see Sect. \ref{sect3} fo the details). The best fit is at $P=18.020\pm0.006\,\mathrm{d}$.}
\label{fsfig} 
\end{figure}

\section{Observations}
Photometric $VI_C$ data were taken with Amadeus, one of the 0.75\,m Automatic Photoelectic Telescopes operated by the Leibniz Institute for
Astrophysics Potsdam at Fairborn Observatory, Arizona \citep{strassmeier2} through 13 September 1999\,--\,27 September 2001 and 15 May 2013\,--\,7 January 2014. All measurements were made differentially with respect to HD\,217019. The check star was HD\,217590. For more details on the reduction process we refer to \cite{strassmeier2} and \cite{granzer1}. In total, we have 365 new observation in $V$ and $I_C$, and 329 points altogether of $V$ and $y$ from Paper I. The mean photometric errors were 0.021 mag. 

\begin{figure*}[t!]
\centering
\includegraphics[width=18cm]{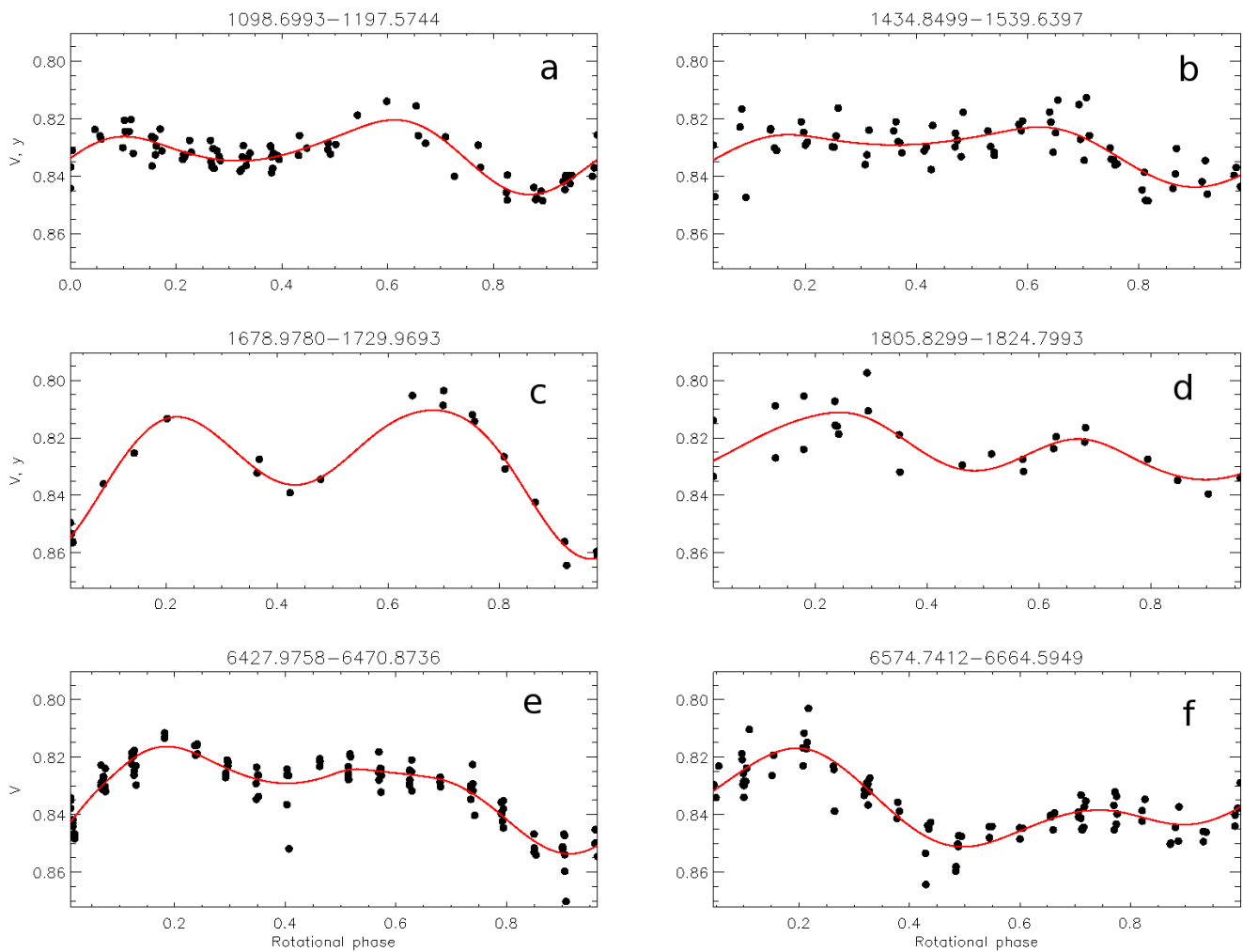}
\caption{$V+y$ (panel {\it a-d}) and $V$ (panel {\it e-f}) data from six different timespans between 8 Oct 1998 and 7 Jan 2014 (corresponding HJDs-2450000 are seen on the top of the panels). Dots are the measurements, the continous line shows the fitted spot model. }
\label{lcfig} 
\end{figure*}

Spectroscopic observations were obtained with the NARVAL high-resolution echelle spectropolarimeter mounted on the 2\,m Bernard Lyot Telescope (TBL) of the Observatoire Midi-Pyrenees at Pic du Midi, France between 1 October and 20 November in 2012. Spectroscopic object mode was used to achieve a peak resolution of $R=80000$. The exposure time $t_{\mathrm{exp}}=600\,\mathrm{s}$ yielded a typical signal-to-noise ratio of 300 at 6400 \AA. Each night, three consecutive spectra were taken and averaged to filter out cosmic beats and to increase signal-to-noise ratio.

The photometric period taken from Paper I is refined (see Sect. \ref{sect3}). Rotational phases were calculated using the following equation:
\begin{equation}\label{eq1}
\mathrm{HJD}=2451639.0033+18.020 \times E.
\end{equation}
For the observing log, see Table \ref{specs}.

Data reduction was carried out with the standard NARVAL pipeline. Wavelength calibration was done by using ThAr arc\-lamps. An additional continuum fitting and normalization was applied to avoid artifacts on the Doppler reconstructions.

\section{Photometric analysis}\label{sect3}

In Paper I, the derived power spectrum from the fast fourier transform (FFT) had two peaks due to the bicentral longitudinal spot distribution. The real rotational period is the longer one, while the shorter (half) period is caused by persistent spots with phase differences of about 0.5. 
From the extended photometric data, we refined the rotational period determination of DI\,Psc. For this, we use the same two methods as in Paper I, i.e. the FFT and string length Lafler  Kinman (SLLK, see \citealt{clarke1}) methods for all available data. The results are plotted in Fig.~ \ref{fsfig}. The most prominent peak from FFT indicates a rotational period of $P_{\mathrm{rot}}=17.84^{+0.31}_{-0.32}\,\mathrm{d}$. This agrees well with $P_{\mathrm{rot}}=18.020\pm0.006\,\mathrm{d}$ derived with the SLLK method (which is more reliable in this case because of the spots at half period) and with the result from Paper I ($P_{\mathrm{rot}}/18.066\pm0.088$). In this paper, we use the refined SLLK period.

\begin{figure*}[t!]
\centering
\includegraphics[width=17cm]{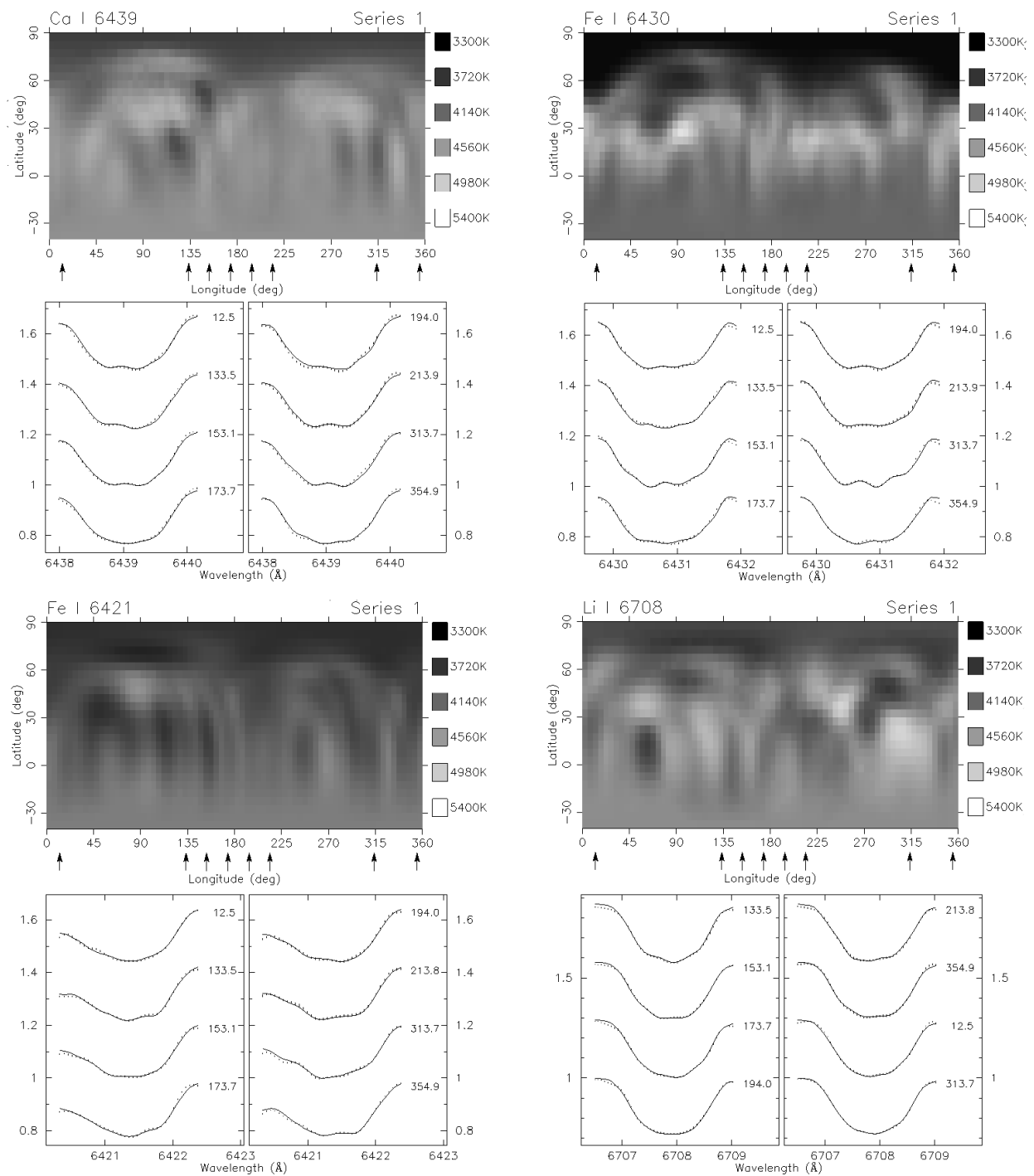}
\caption{Ca\,{\sc i}\,6439 (upper left), Fe\,{\sc i}\,6430 (upper right), Fe\,{\sc i}\,6421 (lower left), and Li\,{\sc i}\,6708 (lower right) maps for the first series of spectra. The temperature maps are presented in pseudo-Mercator projection. The phases of the spectroscopic observations are marked with arrows below the panels. }
\label{difig1} 
\end{figure*}
The new $V+y$ data are shown in Fig. \ref{lcfig} fitted with analytic spot models.
The light curves were inverted with the code \texttt{SpotModel} \citep{sml}. A two-spot model was used, except in panel {\it e}, where a third spot was applied  to obtain an acceptable fit. There are two distinct, well separated active longitudes near  phases 0.0 and 0.5 in every timespan, with slightly changing positions (cf. Paper I).

\section{Doppler imaging}\label{sect4}

\subsection{Doppler imaging code \texttt{TempMap}}
Our Doppler imaging code \texttt{TempMap} (\citealt{rice1}, \citealt{rice2}, \citealt{rice3})  performs a full LTE spectrum synthesis through a series of ATLAS-9 \citep{kurucz1} atmospheric models. Atomic parameters of the transitions were obtained from the VALD database \citep{piskunov1, kupka1}. Local line profiles were calculated for temperatures ranging from 3500\,K to 5500\,K in steps of 250\,K and with solar abundances, except for the Li abundance, where the value A(Li)=2.20 was used. Simultaneous inversions of the usual mapping lines were then carried out using maximum-entropy regularization. 
The astrophysical parameters for the inversion process are given in Table \ref{astdata}.

\begin{figure*}[t!]
\centering
\includegraphics[width=17cm]{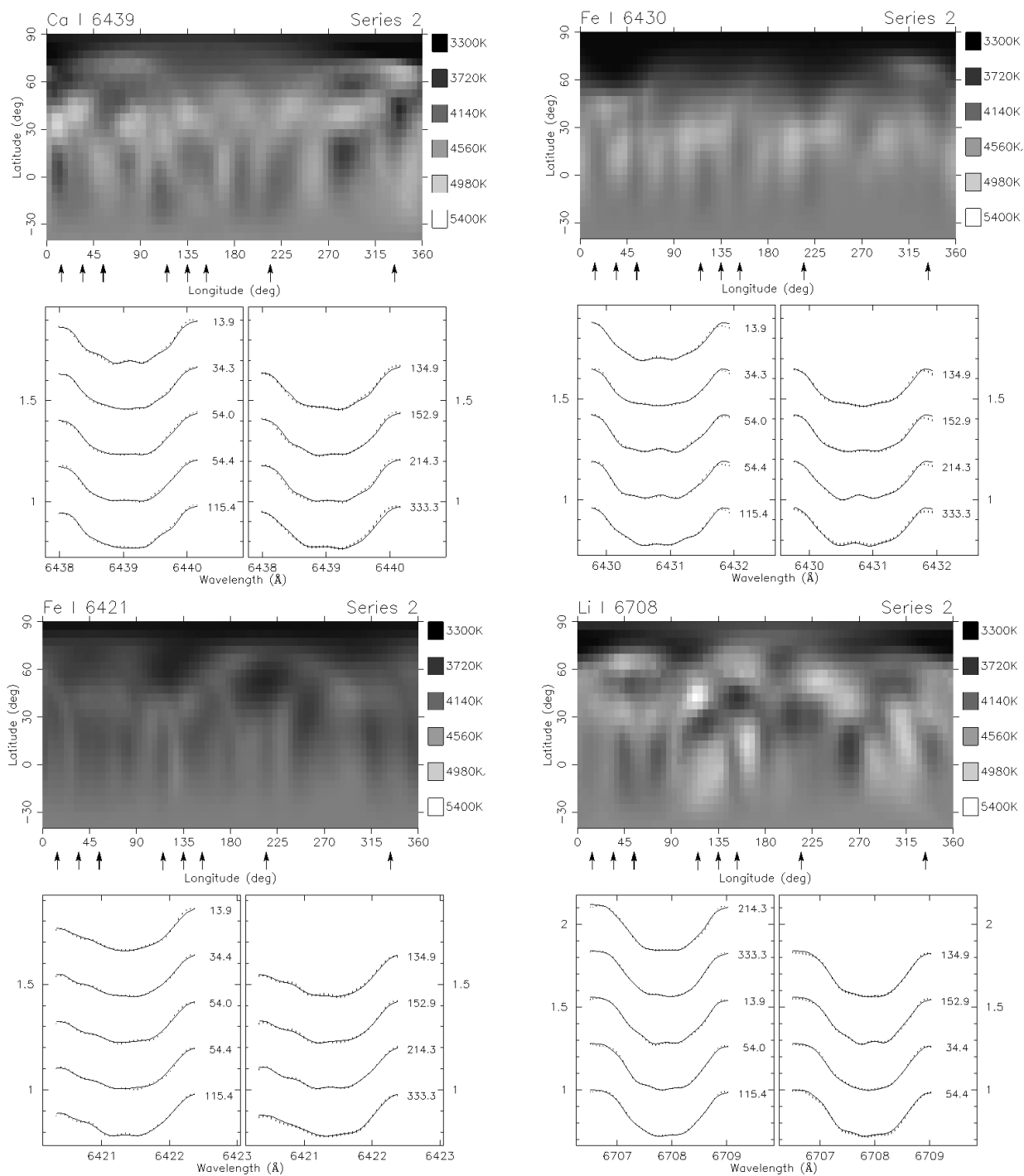}
\caption{Maps for the second series of spectra. For the details, see the caption of Fig. \ref{difig1}.}
\label{difig2} 
\end{figure*}

\subsection{Doppler reconstructions of DI Psc}
The spectroscopic dataset was divided into two subsets, both consisting of eight spectra. The phase coverage is not completely uniform, but is still suitable for deriving subsequent Doppler maps.

Surface reconstructions for the two datasets were carried out for the commonly used inversion lines (Ca\,{\sc i}\,6439, Fe\,{\sc i}\,6430, Fe\,{\sc i}\,6421) of well-known formation physics, as well as for the Li\,{\sc i}\, 6708 line. Resulting maps for the individual lines are plotted in Fig. \ref{difig1} and Fig \ref{difig2}. The maps reveal a cool polar cap, in good agreement with each other, but the temperature ranges differ somewhat, due to the different temperature sensitivity of the mapping lines.
In lower latitudes, spot concentrations are found around $80^{\circ}$ and $300^{\circ}$ longitudes. The two appendages of the polar spot agree with the results of the photometric spot modeling. However, gaps in the phase coverage may affect the reliability of the inversions, or in other words, such arch-like artifacts may appear.

\begin{figure*}[t!]
\centering
\includegraphics[width=18cm]{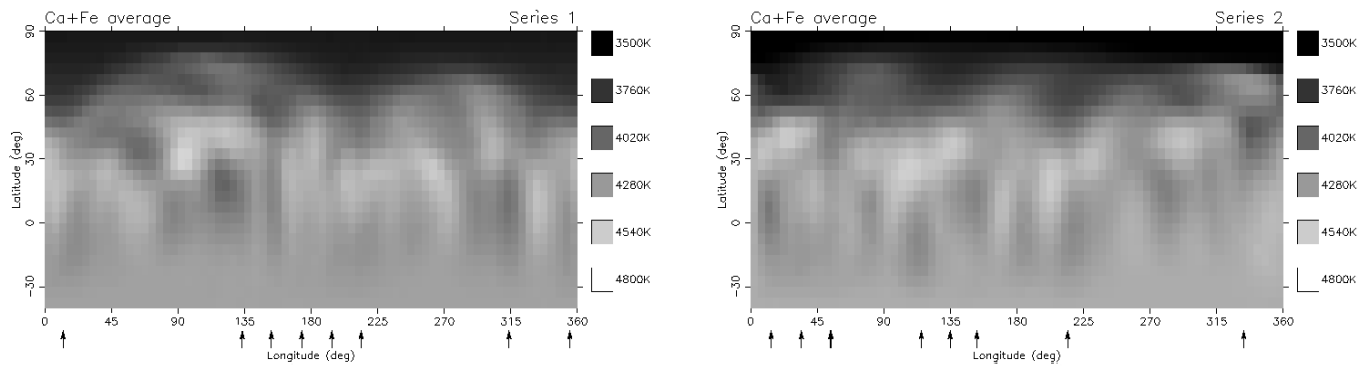}
\caption{Average maps of DI Psc in a pseudo-Mercator projection (Li\,{\sc i}\, 6708 inversions excluded).}
\label{diavg} 
\end{figure*}

In Fig. \ref{diavg}, the combined maps are plotted for the subsequent datasets. The Li maps were excluded from the averaging process and the cross-correlation analysis (see Sect. \ref{sect43}), because of the prominent bright features around $150^{\circ}$ and $300^{\circ}$ that are suspected to be non-activity related phenomena. This issue is further discussed toroughly in Section \ref{sectli}.

\subsection{Surface differential rotation}\label{sect43}
Longitudinal spot displacements from the first series compared to the second can be used as a tracer of surface differential rotation. Such rearrangements might be suspected from the visual comparison of the respective subsequent maps (note the changing shape of the polar cap, or displacement of the features around $300^{\circ}$).

Differential rotation can be measured by cross-correlating consecutive Doppler images \citep{donati1}. We applied our code \texttt{ACCORD} and used average cross-correlation as described e.g. in \cite{kovari2}. 
To fit the cross-correlation pattern, a quadratic rotational law was assumed with the form 
\begin{equation}
\Omega(\beta)=\Omega_{\mathrm{eq}}-\Delta\Omega\sin^2{\beta},
\end{equation} 
where $\Omega_{\mathrm{eq}}$ is the angular velocity of the equator, while $\Delta\Omega=\Omega_{\mathrm{eq}}-\Omega_{\mathrm{pole}}$ defines the difference between the equatorial and polar angular velocities. The dimensonless surface shear coefficient $\alpha$ is defined as $\Delta\Omega/\Omega_{\mathrm{eq}}$.

\begin{figure}[b!!]
\centering
\includegraphics[width=7cm]{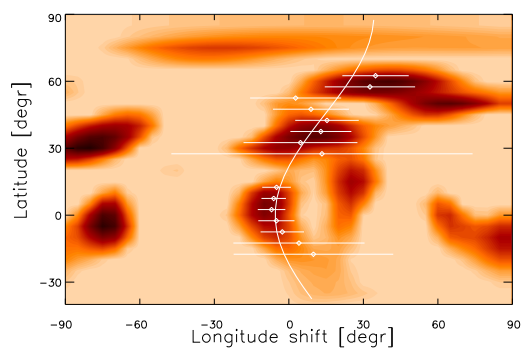}
\caption{Cross-correlation function  map for DI Psc. The image is the average of three correlation maps obtained from cross-correlating our individual Doppler reconstructions (Ca\,{\sc i}\,6439, Fe\,{\sc i}\,6430, Fe\,{\sc i}\,6421), excluding the lithium inversions. The dots with the error bars represents the correlation peaks fitted with Gaussians per latitude bin of $5^{\circ}$. The best fit suggests an antisolar differential rotation of $\alpha=-0.083\pm0.021$.}
\label{ccf} 
\end{figure}

In Fig. \ref{ccf}, the cross-correlation function map is plotted, with the best-fit rotation law to the correlation peaks. Although the cross-correlation pattern is somewhat noisy because we had only two maps per mapping line for comparison, {\bf } it is clear that the near-equatorial features are dropped behind the higher latitude features, indicating an antisolar-type differential rotation at first inspection. On the other hand,  the uncertainty of the inclination may also affect the result. Nevertheless, the cross-correlation was weighted for the temperature contrast  to avoid overinterpretation of the low latitude spots. 

Our fit yields $\Omega_{\mathrm{eq}}=9.70\pm0.15^{\circ}/\mathrm{d}$ and $\Delta\Omega=1.63\pm0.41^{\circ}/\mathrm{d}$, with  $\alpha=-0.083\pm0.021$ surface shear parameter. Errors are estimated from the FWHMs and amplitudes of the Gaussian fits to the latitude bins.

\section{Discussion}

\subsection{Activity cycle length}
The presence of a dominant polar spot in our new maps agree with the results of Paper I, as the low-latitude features does, which were also found in 2000. In 2012, the contrast of the low-latitude spots was somewhat higher, although spot temperatures in general are in the same range,  $T_{\mathrm{eff}}-T_{\mathrm{spot}} \approx 1000 \, \mathrm{K}$.

On the other hand, \cite{lindborg1} found significantly lower spot contrasts of $\approx 500 \, \mathrm{K}$. Their result is consistent with the smaller bumps  in the line profiles of these spectra. \cite{lindborg1} concluded that the spottedness of DI\,Psc is highly erratic, meaning  there was virtually no temperature spot in 2004. One year later, both high- and low-latitude features were detected. Indeed, the spot constrast increased even more in September 2006.

Based on our new Doppler maps of higher spottedness and following \cite{lindborg1}, we estimate that the activity cycle of DI\,Psc is in the range of 5-7 years.

\subsection{Antisolar differential rotation}
\begin{figure}[b!!!]
\centering
\includegraphics[width=8cm]{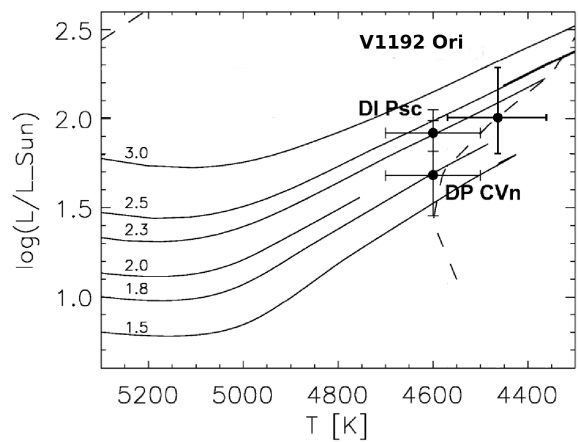}
\caption{Evolutionary tracks taken from Paper I, overplotted with the most likely positions of DI\,Psc, DP\,CVn and V1192\,Ori. The dashed lines indicate the borders of the first dredge-up phase.}
\label{hrd_li} 
\end{figure}

In Paper I, antisolar differential rotation was also detected on DP\,CVn, a "twin" of DI\,Psc. Furthermore, an antisolar-type rotational law was reported for V1192\,Ori (HD\,31993), a K2III single giant (\citealt{strassmeier3, weber1}). According to \cite{fekel1}, V1192\,Ori exhibits a lithium abundance of $A(\mathrm{Li})=1.4\pm0.2$, which indicates that the star is around the end of its first dregde-up phase, similarly to DI\,Psc and DP\,CVn (cf. Paper I and \citealt{strassmeier3}).
Fig. \ref{hrd_li} shows the three stars on the H-R diagram.
We also note that DI\,Psc, DP\,CVn and V1192\,Ori are currently the only known single K giants with detected antisolar-type differential rotation, and all of them are at the end of their first dredge-up. This may indicate 
that a process occurs at about this very position of the RGB, which can account for both the antisolar direction of the differential rotation and the lithium-enhancement through a stronger meridional circulation capable of transporting Li-rich material from the inner parts towards the surface (and the poles) along with angular momentum (\citealt{fekel1}, \citealt{kitchatinov1}).

\subsection{Investigating the surface Li-enchancement -- signs of lithium enchancement that is not related to activity}\label{sectli}
From the visual inspection of the indvidual Doppler maps, it is clear that the Li maps significantly  differ from maps of the other mapping lines, while the two Li maps have several similar features (note the bright features around $150^{\circ}$ and $300^{\circ}$). These hot spots are probably not caused by insufficient phase coverage, because there are measurements around the given phases, and the phase distribution for the subsequent maps are different. Lithium spots are probably related more to chemical inhomogeneities. On the other hand, features appearing at both rotations could be more persistent than (regular) spots. This might indicate that the Li-enrichment is caused by processes that are not related to activity.
\begin{figure*}[t]
\centering
\includegraphics[width=19cm]{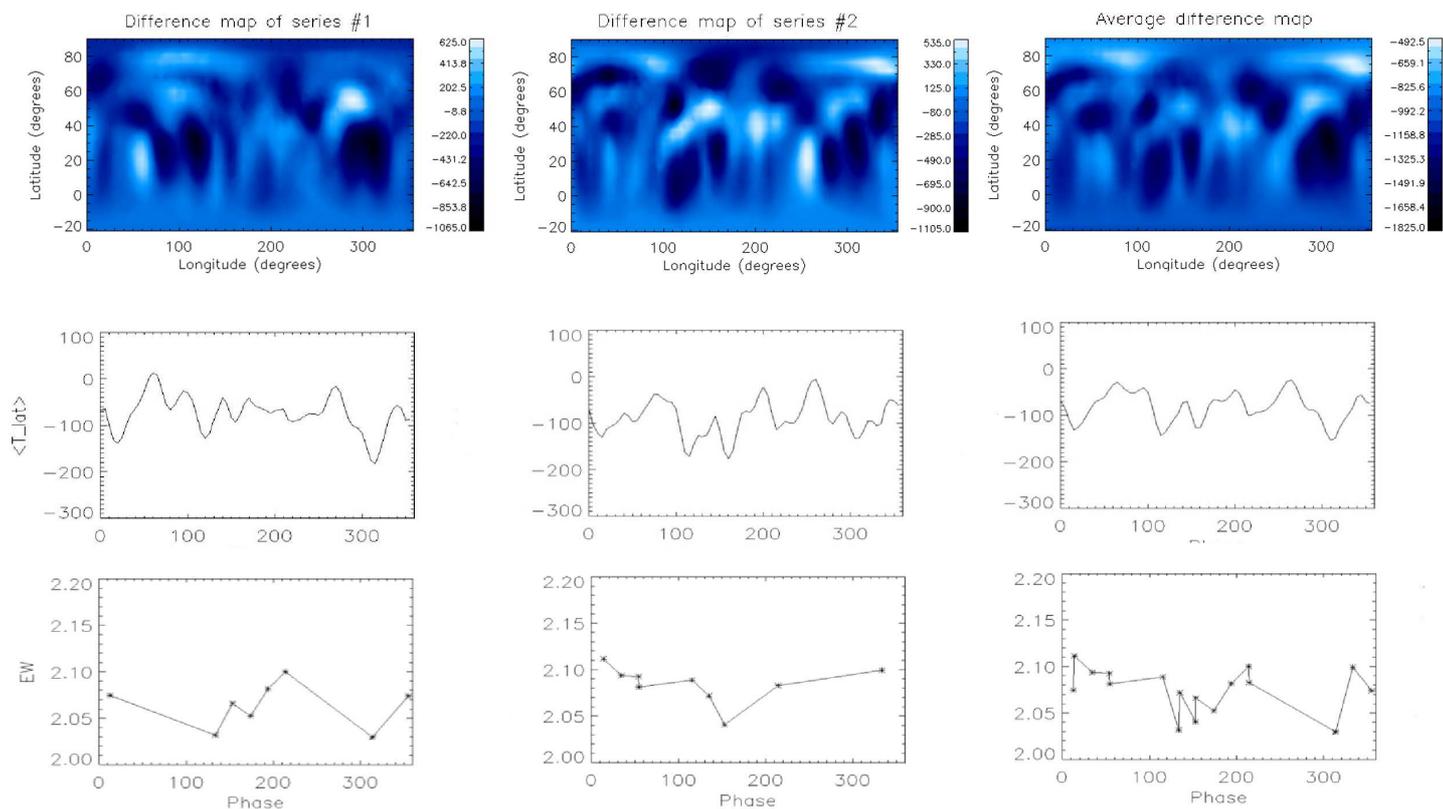}
\caption{Top panels: difference maps by substracting the Li-maps from the corresponding Ca+Fe average maps of the subsequent series (plotted in Fig \ref{diavg}). In the last column, plotted are the averaged difference maps. Note the similar high-latitude features at $\approx0^{\circ}$ and $\approx230^{\circ}$ and the low-latitude ones around $150^{\circ}$ and $300^{\circ}$. Middle panels: the difference maps are averaged latitudinally. Bottom panels: measured equivalent widths of the Li\,{\sc i}\,6708 line.}
\label{diff} 
\end{figure*}
Our temperature inversion code assumes constant element abundances over the surface. Therefore, a bump on the line profile caused by a chemical abundance enhancement would be interpreted as a hotter region, as shown in the maps at $\approx150^{\circ}$ and $\approx300^{\circ}$. 
To investigate the surface lithium distribution, we performed an indirect test by subtracting the Li maps from the corresponding averaged Ca+Fe images. This way, the hotter spots will presumably appear on the difference maps as negative temperature differences, indicating lithium abundance spots.

The difference maps are shown in the top row of Fig. \ref{diff}. Note that the spots of negative pixel values (i.e. dark features, indicating suspected Li abundance spots) around $150^{\circ}$ and $300^{\circ}$ at low latitudes, and the high latitude features at $\approx0^{\circ}$ and $\approx220^{\circ}$ are apparent on the difference maps of both series, meaning they are stable. In addition, the shape and structure of the low-latitude features around $150^{\circ}$ and $300^{\circ}$ seem to be persistent. The possibility of an artificial origin can be ruled out, since inadequate phase-coverage would cause arch-like structures, which are not seen. 
Moreover, the spots of negative pixels can be seen at the same phases, but the phase distribution of the observations is different in the subsequent maps. The heavy spottedness at lower latitudes might bias this finding however.

The polar spot in the difference maps may indicate an enhanced lithium abundance around the pole. This would agree with the antisolar differential rotation, considering that lithium-rich material can be transported towards the surface by an enhanced meridional flow, which would elevate the freshly synthesized Li to the outer regions and simultaneously move it towards the pole. Along with the lithium, material of higher angular momentum is also transported and carried to the polar regions, causing antisolar type differential rotation (cf. \citealt{kitchatinov1}).

\subsection{Rotational modulation of the Li equivalent widths}
We determine the rotational modulation of the measured Li\,{\sc i}\,6708 line equivalent widths (hereafter EWs). The result is plotted in the bottom panels of Fig. \ref{diff}. In the middle panels, the difference maps are averaged latitudinally.
Visual inspection of the two curves reveals that lower temperature values coincide with lower lithium EWs. However, a correlation analysis does not confirm this, yielding $r=0.17$. Nevertheless, the rotational modulation of the EW suggests that one single measurement of the Li line may not be sufficient to determine precise abundance values (cf. \citealt{fekel2}).

\section{Summary and conclusions}
We applied Doppler imaging for high-resolution time-series spectroscopic data in order to reconstruct the temperature distribution, surface differential rotation, and investigate the possible imhomogenities in the surface lithium enrichment. 

Our conclusions are as follows:

\begin{itemize}
\item Photometric spot modeling at six different time intervals shows two active longitudes slightly moving in phase, which is constistent with the light curves and Doppler maps in Paper I. This finding is also supported by our new Doppler reconstructions. \\

\item Doppler maps of two subsequent datasets were reconstructed for the Ca\,{\sc i}\,6439, Fe\,{\sc i}\,6430, Fe\,{\sc i}\,6421, and Li\,{\sc i}\,6708 mapping lines. The reconstructions reveal prominent polar spots, and various low-latitude features. Individual Ca and Fe maps agree well, but Li maps significantly differ, for example by showing bright features at lower latitudes. \\

\item Li maps indicate surface abundance inhomogenities, which may be related to non-uniform Li-enchancement	.\\

\item Our cross-correlation technique yields an antisolar-type differential rotation of $\alpha=-0.083\pm0.021$ surface shear coefficient. This  finding resembles the differential rotational pattern of DP\,CVn and V1192\,Ori, two lithium-rich giants. We note, however, that more observations are necessary to confirm this result. \\

\item Based on the comparison of our new results with Paper I and \cite{lindborg1}, we estimate that the activity cycle of DI\,Psc is in the range of 5-7 years. 
\end{itemize}

\small{
\emph{
Acknowledgements.} The authors would like to thank the anonymous referee for his/her useful insights. LK, ZsK, KV, and OK are grateful to the Hungarian Science Research Program (OTKA) for support under the grant K-81421 and K-109276. This work is supported by the "Lend\"ulet-2009" and "Lend\"ulet-2012" Young Researchers' Programs of the Hungarian Academy of Sciences. The data used for this paper were gathered via OPTICON FP7. 
}


\end{document}